\documentclass[useAMS,usenatbib]{mn2e}

\usepackage[dvips]{graphicx}

\title[Magnetorotational supernovae]{Magnetorotational supernovae}
\author[N.V.Ardeljan, G.S.Bisnovatyi-Kogan, and S.G.Moiseenko]
{N.V.Ardeljan$^{1}$\thanks{E-mail: ardel@cs.msu.su (NVA)},
G.S.Bisnovatyi-Kogan$^{2}$\thanks{gkogan@iki.rssi.ru (GSBK)} and
S.G.Moiseenko$^{2}$\thanks{moiseenko@iki.rssi.ru (SGM)}\\
$^{1}$Department of Computational Mathematics and
Cybernetics, Moscow State University, Vorobjevy Gory, Moscow B-234, Russia\\
$^{2}$Space Research Institute, Profsoyuznaya str. 84/32, Moscow 117997,
Russia}
\begin{document}

\date{Accepted  . Received ; in original form }

\pagerange{\pageref{firstpage}--\pageref{lastpage}} \pubyear{2003}

\maketitle

\label{firstpage}

\begin{abstract}
We present the results of 2D simulations of the magnetorotational model of a supernova
explosion. After the core collapse the core consists of rapidly a rotating
proto-neutron star and a differentially rotating envelope.
The  toroidal part of the magnetic energy generated by the differential
rotation grows linearly with time at the initial stage of the evolution of
the magnetic field. The linear growth of the toroidal magnetic field is
terminated by the development of magnetohydrodynamic instability, leading
to drastic acceleration in the growth of magnetic energy. At the moment
when the magnetic pressure becomes comparable with the gas pressure at the
periphery of the proto-neutron star $\sim 10-15$km from the star centre the
MHD compression wave appears and goes through the envelope of the collapsed
iron core. It transforms soon to the fast MHD shock and produces a supernova
explosion. Our simulations give the energy of the explosion $0.6\cdot 10^{51}$ ergs.
The amount of the mass ejected by the explosion is $\sim 0.14M_\odot$.
The implicit numerical method, based on the Lagrangian triangular grid of
variable structure, was used for the simulations.

\end{abstract}


\section{Introduction}

The problem of explaining core collapse supernova explosions is one of
the long standing and fascinating problems in astrophysics. The most
popular core collapse explosion mechanisms up to now were the prompt
explosion caused by the bounce shock and the
 neutrino driven mechanism. The 1D spherically symmetrical simulations of the core
collapse supernova do not lead to the explosion (see for example {\citealt{Burrows1995},
\citealt{Buras2003} and references therein). The 2D and 3D simulations of the neutrino driven
supernova mechanism do not give supernova explosions either with a sufficient level of confidence.
The recently improved models of the core collapse where the neutrino transport was
simulated by solving the Boltzmann equation do not explode either (\citealt{Buras2003}).

The important role of the rotation and the magnetic fields for the core collapse supernova
was suggested  by \citet{bk1970}. The idea of the magnetorotational mechanism
consists of angular momentum transfer outwards using a magnetic field, which twists due
to the differential rotation. The toroidal component of the magnetic field amplifies
with time, it leads an increase in magnetic pressure and generation of a
supernova shock.

The first 2D simulations of the collapse of the rotating magnetized star were
presented in the paper of \citet{leblanck}, with unrealistically large values of the
magnetic field. The differential rotation and amplification of
the magnetic field resulted in the formation of the axial jet.
A semi-qualitative picture of the magnetorotational explosions was considered by \citet{meier}.
The results of
2D simulations of the similar problem were given in by \citet{ohnishi} and
\citet{symbalisty}. Recently, simulations of the
magnetorotational processes in the collapsed core have been represented
by \citet{kotake} and \citet{yamada}, where the authors performed 2D simulations of
the magnetorotational processes for a strongly magnetized core collapse.
They assumed that the initial poloidal magnetic field is uniform and parallel
to the rotation axis. In \citet{kotake} the initial strong toroidal magnetic field was also used as
the initial condition. In their case the toroidal magnetic field was not equal to zero at the rotational axis $z$,
which means a formally nonphysical situation with infinite current density at the rotational axis.
The values of the magnetic fields used by them are usual for "magnetars" but too large for the
ordinary  core collapse supernova.
1D simulations of the successful magnetorotational supernova explosion for a wide range of the initial
values of the magnetic field are represented by \citet{abkp}.
2D simulations of the magnetorotational explosion in a rotating magnetized
cloud have been described by
\citet{abkm}.

One of the main problems for the numerical simulation of the magnetorotational
supernova is the smallness of the initial magnetic fields. The ratio of the
initial magnetic and gravitational energies is in the range of $10^{-6}-10^{-10}$.
This value characterizes the stiffness of the set of MHD equations. We have a
problem with  strongly varying characteristic timescales. The small one is
defined by a very high sound speed in the center of the core and the large one
is a characteristic time of the evolution of the magnetic field. In such a situation the explicit
numerical methods which are widely used in astrophysical hydro-simulations require an enormously
large number of timesteps and a possible loss of accuracy due to the large numerical errors.
The implicit approach should be used in this case. It is well known that implicit schemes are
free from the Courant timestep limitation, which is too strong for such problems.

At the initial stages of the process the toroidal magnetic field is produced by a twisting of the radial
component due to differential rotation and is growing linearly, $B_\varphi \sim B_r \omega t$.
The linear growth is terminated when the toroidal component becomes so strong that magnetohydrodynamic
instability starts to develop. This instability leads to poloidal motion of the matter,
which increases the radial component,  which
in turn strongly amplifies the growth of the toroidal field. As a result,
both components, toroidal and poloidal, start to grow almost exponentially, shortening greatly the time
between the collapse and magnetorotational explosion.

The differential rotation and amplification of the toroidal component of the magnetic field leads to
the formation of the MHD compression wave moving from the center of the star. This compression
wave goes along a steeply decreasing density profile and soon transforms into the MHD shock wave.
This MHD shock is supported by the flux of the matter from the central parts of the star.
This matter flux works like a piston increasing the velocity and the strength of the shock.
When the MHD shock reaches the outer boundary of the collapsed iron core, the kinetic energy of the
radial motion (supernova explosion energy) is  $\sim0.6 \cdot 10^{51}$ergs. The amount of mass
ejected by the shock is about $0.14 M_\odot$ of the total mass of the star, $1.2 M\odot$.

\section{Equations of state and neutrino losses}
For the simulations we use the equation of state used in \citet{abkpch}:

\begin{eqnarray*}
  P \equiv P(\rho,T)=P_0(\rho)+\rho \Re T + \frac {\sigma T^4} {3},
\end{eqnarray*}

\begin{equation}\label{pressure}
P_0(\rho)=\left\{
\begin{array}{l}
P_0^{(1)}=b_1\rho^{5/3}/(1+c_1\rho^{1/3}),0 \leq\rho\leq\rho_1,\\
P_0^{(k)}=a\cdot 10^{b_k({\textrm{lg}}\rho-8.419)^{c_k}},\rho_{k-1}\leq\rho\leq\rho_k,\\
\hskip 4cm k=\overline{2,6}\\
\end{array}
\right.
\end{equation}
\[
\begin{array}{lll}
b_1=10^{12.40483}   & c_1=10^{-2.257} & \rho_1=10^{9.419}    \\
b_2=1.           & c_2=1.1598      & \rho_2=10^{11.5519}  \\
b_3=2.5032       & c_3=0.356293    & \rho_3=10^{12.26939} \\
b_4=0.70401515   & c_4=2.117802    & \rho_4=10^{14.302}   \\
b_5=0.16445926   & c_5=1.237985    & \rho_5=10^{15.0388}  \\
b_6=0.86746415   & c_6=1.237985    & \rho_6\gg \rho_5     \\
a=10^{26.1673},  &                 &
\end{array}
\]
where $\Re$ is the gas constant equal to $0.83\cdot 10^8 \frac
{\mathrm {cm^2}}{\mathrm s^2}$K, $\sigma$ is the constant of the radiation density,
$P$ is pressure, $\rho$ is density, and $T$ is temperature. In the
expression $P_0(\rho)$ the value $\rho$ was identified with the
total mass-energy density. For the cold degenerated matter
expression for $P_0(\rho)$ is the approximation of the tables from
\citet{bps, mjb}.

In the neighborhood of points $\rho=\rho_k$ in formula (\ref{pressure}) the function
$P_0(\rho)$ was smoothed in the same way as in \citet{abkpch}, to make
continuous the derivative
 ${\textrm d}P_0/{\textrm d} \rho$ :

\begin{equation}\label{smoothp}
P_0(\rho)=\left\{
\begin{array}{l}
P_0^{(k)}, \rho\in[\rho_{k-1}+\xi_{k-1},\rho_k-\xi_k],\\
\hskip 1cm k=\overline{1,6}, \> \rho_0+\xi_0=0,\\
\theta_k P_0^{(k)}+(1-\theta_k P_0^{k+1}),\\
\hskip 1cm \rho\in[\rho_k-\xi_k,\rho_k+\xi_k],  k=\overline{1,5},\\
\end{array}
\right.
\end{equation}
where
$$
\theta_k=\theta(\rho)=\frac{1}{2}-\frac{1}{2}\sin\left(\frac{\pi}{2\xi_k}(\rho-\rho_k)\right), \quad
\xi_k=0.01\rho_k.
$$

The specific energy (per mass unit) was defined thermodynamically as:
\begin{equation}\label{inten}
\varepsilon=\varepsilon_0(\rho)+\frac{3}{2}
\Re T +\frac{\sigma T^4}{\rho}+\varepsilon_{Fe}(\rho,T).
\end{equation}

The value $\varepsilon_0(\rho)$ is defined by the relation
\begin{equation}\label{inten0}
\varepsilon_0(\rho)=\int\limits_0^\rho \frac{P_0(\tilde{\rho})}{\tilde{\rho}^2}\textrm{d}\tilde{\rho}.
\end{equation}

 The term  $\varepsilon_{Fe}$ in equation (\ref{inten}) is responsible for iron dissociation.
 It is used in the following form:
\begin{equation}\label{ferrum0}
  \varepsilon_{Fe}(\rho,T)=\frac{E_{b,Fe}}{A\>m_p}
  \left(\frac{T-T_{0Fe}}{T_{1Fe}-T_{0Fe}}\right).
\end{equation}
It was supposed that in  region of the iron dissociation
the iron amount was about $50\%$ of the mass,
$E_{b,Fe}=8\cdot 10^{-5}$erg is the iron binding energy,
$A=56$ is the iron atomic weight, and $m_p=1.67\cdot 10^{-24}$g is the proton mass,
$T_{0Fe}=0.9\cdot 10^{10}$K, $T_{1Fe}=1.1\cdot 10^{10}$K.
For the numerical calculations formula (\ref{ferrum0}) has been
slightly modified (smoothed):
\begin{equation}\label{ferrum1}
  \varepsilon_{Fe}(\rho,T)=\frac{E_{b,Fe}}{A\>m_p}
 \frac{\left(1+\sin\left(\pi  \left(\frac{T-T_{0Fe}}{T_{1Fe}-T_{0Fe}}\right)-\frac{\pi}{2}\right)\right)}{2}.
\end{equation}

The neutrino losses for Urca processes are used in the form, taken from \citet{bkpopsam},
approximating the table of \citet{iin}:
\begin{equation}\label{urca}
  f(\rho,T)=\frac {1.3 \cdot 10^9 {\textrm {\ae}}(\overline{T})\overline{T}^6}
  {1+(7.1\cdot 10^{-5}\rho \overline{T})^{\frac{2}{5}}}\quad
  {\textrm {erg}} \cdot {\textrm{g}}^{-1} \cdot {\textrm {s}}^{-1},
\end{equation}

\begin{equation}
{\textrm {\ae(T)}}=\left\{
\begin{array}{rclcccccc}
1,&\overline{T}<7,\\
664.31+51.024 (\overline{T}-20), & 7\leq \overline{T} \leq 20,\\
664.31, & \overline{T}>20,
\end{array}
\right.
\end{equation}

$$\overline {T}=T\cdot 10^{-9}.$$

The neutrino losses from pair annihilation, photo production, and plasma
were also taken into account. These types of the neutrino losses have been
approximated by the interpolation formulae
from \citet{schindler}:
\begin{equation}\label{dopneu}
  Q_{\mathrm tot}=Q_{\mathrm pair}+Q_{\mathrm photo}+Q_{\mathrm plasm}.
\end{equation}
The three terms in  (\ref{dopneu}) can be written in the following general form:
\begin{equation}\label{schindler1}
  Q_d=K(\rho,\alpha)e^{-c\xi}\frac{a_0+a_1\xi+a_2\xi^2}{\xi^3+b_1\alpha+b_2\alpha^2+b_3\alpha^3}.
\end{equation}
For $d=pair$, $K(\rho,\alpha)=g(\alpha)e^{-2\alpha}$,
$$
g(\alpha)=1-\frac{13.04}{\alpha^2}+\frac{133.5}{\alpha^4}+\frac{1534}{\alpha^6}+\frac{918.6}{\alpha^8};
$$
For $d=photo$, $K(\rho,\alpha)=(\rho/\mu_Z)\alpha^{-5};$\\
For $d=plasm$, $K(\rho,\alpha)=(\rho/\mu_Z)^3;$
$$
\xi=\left(\frac{\rho / \mu_Z}{10^9}\right)^{1/3} \alpha.
$$

Here, $\mu_Z=2$ is the number of nucleons per electron. Coefficients $c,\> a_i,$ and $b_i$
for the different losses are given in the Table \ref{tabb1} from \citet{schindler}.
\begin{table*}
 \centering
 \begin{minipage}{140mm}
  \caption{Coefficients for formula (\ref{schindler1}) from \citet{schindler}.}\label{tabb1}
   \begin{tabular}{|l|l|l|l|l|l|l|l|}
  \hline
       & $a_0$ & $a_1$ & $a_2$ & $b_1$ & $b_2$ & $b_3$ & $c$ \\ \hline
  \multicolumn{8}{|c|}{$10^8\>{\textrm K} \leq T \leq \> 10^{10}\>{\textrm K}$}\\ \hline
  pair  & 5.026(19) & 1.745(20) & 1.568(21) & 9.383(-1) & -4.141(-1) & 5.829(-2) & 5.5924 \\
  photo & 3.897(10) & 5.906(10) & 4.693(10) & 6.290(-3) & 7.483(-3) & 3.061(-4) & 1.5654 \\
  plasm & 2.146(-7) & 7.814(-8) & 1.653(-8) & 2.581(-2) & 1.734(-2) & 6.990(-4) & 0.56457 \\ \hline
  \multicolumn{8}{|c|}{$10^{10}\>{\textrm K} \leq T \leq \> 10^{11}\>{\textrm K}$}\\ \hline
  pair  & 5.026(19) & 1.745(20) & 1.568(21) & 1.2383 & -8.1141(-1) & 0.0 & 4.9924 \\
  photo & 3.897(10) & 5.906(10) & 4.693(10) & 6.290(-3) & 7.483(-3) & 3.061(-4) & 1.5654 \\
  plasm & 2.146(-7) & 7.814(-8) & 1.653(-8) & 2.581(-2) & 1.734(-2) & 6.990(-4) & 0.56457 \\ \hline
\end{tabular}
\end{minipage}
\end{table*}
The general formula for the neutrino losses in a nontransparent star has been written in the
form, used by \citet{abkkm2004}:
\begin{equation}\label{neuttot}
  F(\rho,T)=(f(\rho,T)+Q_{tot})e^{-\frac{\tau_\nu}{10}}.
\end{equation}
The multiplier $e^{-\frac{\tau_\nu}{10}}$
in formula (\ref{neuttot}), where $\tau_\nu=S_\nu n l_\nu$,
restricts the neutrino flux for non zero depth to neutrino interaction
with matter $\tau_\nu$. The cross-section for this interaction $S_\nu$ was presented in the form:
$$S_\nu=\frac{10^{-44}T^2}{(0.5965 \cdot 10^{10})^2},$$
the concentration of nucleons  is
$$n=\frac{\rho}{m_p}.$$
The characteristic length scale $l_\nu$, which defines the depth for the neutrino absorbtion, was taken
to be equal to the characteristic length of the density variation as:
\begin{equation}\label{depth}
l_\nu=\frac{\rho}{
|\nabla \rho|}=\frac{ \rho}{\left((\partial \rho /\partial r)^2+
(\partial \rho / \partial z)^2\right)^{1/2}}.
\end{equation}
The value $l_\nu$ monotonically decreases when moving to the outward boundary; its maximum
is in the center. It approximately determines the depth of the neutrinoabsorbing matter.
The multiplier $1/10$ in the expression $e^{-\tau_\nu\frac{1}{10}}$ was applied
because in the degenerate matter of the hot neutron star only some of the nucleons with the
energy near Fermi boundary, approximately $1/10$, take part in the neutrino processes.

\section{Basic equations}
Consider a set of magnetohydrodynamical equations with
self\-gra\-vi\-ta\-tion and infinite conductivity:
\begin{eqnarray}
\frac{{\rm d} {\bf x}} {{\rm d} t} = {\bf v}, \nonumber \\
\frac{{\rm d} \rho} {{\rm d} t} +
\rho \nabla \cdot {\bf v} = 0,  \nonumber\\
\rho \frac{{\rm d} {\bf v}}{{\rm d} t} =-{\rm grad}
\left(P+\frac{{\bf H} \cdot {\bf H}}{8\pi}\right) +
\frac {\nabla \cdot({\bf H} \otimes {\bf H})}{4\pi} -
\rho  \nabla \Phi, \nonumber\\
\rho \frac{{\rm d}}{{\rm d} t} \left(\frac{{\bf H}}{\rho}\right)
={\bf H} \cdot \nabla {\bf v},\>
\Delta \Phi=4 \pi G \rho,
\label{magmain}\\
\rho \frac{{\rm d} \varepsilon}{{\rm d} t} +P \nabla \cdot {\bf v}+\rho F(\rho,T)=0,
 \nonumber\\
P=P(\rho,T),\> \varepsilon=\varepsilon(\rho,T), \nonumber
\end{eqnarray}
where $\frac {\rm d} {{\rm d} t} = \frac {\partial} {
\partial t} + {\bf v} \cdot \nabla$ is the total time
derivative, ${\bf x} = (r,\varphi , z)$, ${\bf v}=(v_r,v_\varphi,v_z)$ is the velocity
vector, $\rho$ is the density, $P$ is the pressure,  ${\bf
H}=(H_r,\> H_\varphi,\> H_z)$ is the magnetic field vector, $\Phi$ is
the gravitational potential, $\varepsilon$ is the internal energy, $G$ is
gravitational constant, ${\bf H} \otimes {\bf H}$ is the tensor
of rank 2, and
$F(\rho,T)$ is the rate of neutrino losses.

$r$, $\varphi$, and  $z$ are spatial Lagrangian coordinates, i.e. $r=r(r_0,\varphi_0,$ and $z_0,t)$,
$\varphi=\varphi(r_0,\varphi_0,z_0,t)$, and $z=z(r_0,\varphi_0,z_0,t)$, where $r_0,\varphi_0,z_0$ are the
initial coordinates of material points of the matter.

Taking into account symmetry assumptions
($ \frac \partial {\partial \varphi} = 0$),
the divergency of the
tensor ${\bf H} \otimes {\bf H}$ can be presented in the
following form:
$$
{\rm \nabla\cdot}({\bf H} \otimes {\bf H})=
\left(\begin{array}{l}
\frac {1}{r} \frac {\partial(rH_rH_r)}{\partial r} +
\frac {\partial(H_zH_r)} {\partial z}-
\frac {1}{r} H_\varphi H_\varphi \\
\frac {1}{r} \frac {\partial(rH_rH_\varphi)}{\partial r} +
\frac {\partial(H_zH_\varphi)} {\partial z}+
\frac {1}{r} H_\varphi H_r \\
\frac {1}{r} \frac {\partial(rH_rH_z)}{\partial r} +
\frac {\partial(H_zH_z)} {\partial z}
\end{array}
\right).
$$

Axial symmetry ($\frac \partial {\partial
\varphi}=0$, $r\geq 0$) and symmetry to the equatorial plane ($z\geq0$) are
assumed. The problem is solved in the restricted domain. At $t=0$ the domain is restricted
by the rotational axis $r\geq 0$, equatorial plane $z\geq 0$, and outer boundary of the star where
 the density of the matter is zero, while poloidal
components of the magnetic field $H_r$, and $H_z$ can be non-zero.

We assume axial and equatorial symmetry ($r\geq 0,\> z\geq 0$).
At the rotational axis ($r=0$) the following boundary conditions are defined: $(\nabla \Phi)_r=0,\> v_r=0$.
At the equatorial plane ($z=0$) the boundary conditions are: $(\nabla \Phi)_z=0,\> v_z=0$.
At the outer boundary (boundary with vacuum) the following condition  is defined:
 $P_{\textrm {outer boundary}}=0$.

We avoid explicit calculations of the function $\varepsilon_0(\rho)$ in (\ref{inten0}), because
this term is eliminated from (\ref{magmain}) due to adiabatic equality:
\begin{equation}\label{adiab}
\rho\frac{{\rm d} \varepsilon_0}{{\rm d} t}=
-\frac{P_0}{\rho}\frac{{\rm d}\rho}{{\rm d} t}=
P_0 \nabla \cdot {\bf v},
\end{equation}
determining the fully degenerate part of EOS. Therefore, defining
\begin{eqnarray*}
  \varepsilon^*=\frac{3}{2}\Re T +\frac{\sigma T^4}{\rho}+\varepsilon_{Fe}(\rho,T),\\
  P^*=\rho \Re T +\frac{\sigma T^4}{3}.
\end{eqnarray*}
The equation for the internal energy in (\ref{magmain}) can be written in
the following form:

\begin{equation}\label{eqmod}
\rho\frac{{\rm d} \varepsilon^*}{{\rm d} t}+P^*
\nabla\cdot{\bf v}+\rho F(\rho,T)=0.
\end{equation}

\section{Dimensionless form of equations}
For the simulations we rewrite the set of equations (\ref{magmain}) in the dimensionless form.
The basic scale values are:
\begin{eqnarray}
r_0=1.35\cdot10^8\textrm{cm},\>\rho_0=10^9{\textrm{g/cm}}^3,\nonumber\\
G=6.67\cdot 10^{-8} {\textrm{cm}^3}/({\textrm {g$\cdot$s}^2}). \label{scales}
\end{eqnarray}
The dimensional functions can be represented in the following
way (values with $\tilde{\phantom a}$ tilde sign are dimensionless values):
\begin{eqnarray}
r=\tilde{r}r_0,\> z=\tilde{z}r_0,\> \rho=\tilde{\rho}\rho_0,\>
v=\tilde{v}v_0,  \nonumber \\ \label{obezraz}
t=\tilde{t}t_0,\>
v_r=\tilde{v}_r v_0,\>
v_\varphi=\tilde{v}_\varphi v_0,\>
v_z=\tilde{v}_zv_0,\>
p=\tilde{p}p_0,\\
T=\tilde{T}T_0,
\Phi=\tilde{\Phi}\Phi_0=\tilde{\Phi}4\pi G\rho_0r_0^2,\>
\varepsilon=\tilde{\varepsilon}\varepsilon_0, \> H=\tilde{H}H_0,
\end{eqnarray}
where
\begin{eqnarray}
 v_0=\sqrt{4\pi G \rho_0 r_0^2}=3.908\cdot10^9\textrm{cm/s},\nonumber\\
 t_0=\frac{r_0}{v_0},\>
 p_0=\rho_0v_0^2,\>
 T_0=\frac{v_0^2}{\Re},\nonumber \\
 \Phi_0=4\pi G\rho_0r_0^2,\> \varepsilon_0=v_0^2,\> H_0=\sqrt{p_0}=x_0t_0^{-1}\rho_0^{-1}. \nonumber
\end{eqnarray}
Taking into account (\ref{eqmod}) the set of basic equations (\ref{magmain}) can be written
in the following dimensionless form
(the tilde sign $\tilde{\phantom a}$ being omitted here):
\begin{eqnarray}
\frac{{\rm d} {\bf x}} {{\rm d} t} = {\bf v}, \nonumber \\
\frac{{\rm d} \rho} {{\rm d} t} +
\rho \nabla \cdot {\bf v} = 0,  \nonumber\\
\rho \frac{{\rm d} {\bf v}}{{\rm d} t} =-{\rm grad}
\left(P+\frac{{\bf H} \cdot {\bf H}}{8\pi}\right) +
\frac {\nabla \cdot({\bf H} \otimes {\bf H})}{4\pi} -
\rho  \nabla \Phi, \nonumber\\
\rho \frac{{\rm d}}{{\rm d} t} \left(\frac{{\bf H}}{\rho}\right)
={\bf H} \cdot \nabla {\bf v},\>
\Delta \Phi= \rho,
\label{eqob}\\
\rho \frac{{\rm d} \varepsilon^*}{{\rm d} t} +P^* \nabla \cdot {\bf v}+\rho F(\rho,T)=0,
 \nonumber\\
P=P(\rho,T),\> \varepsilon=\varepsilon(\rho,T), \nonumber
\end{eqnarray}
\section{Initial conditions}
We have used the model I from \citet{abkpch} as initial conditions . At first we calculated
spherically symmetrical stationary model with central density
$\rho_c=4.5\cdot10^9 \textrm {g/cm}^3$.
The value of the central density corresponds to the maximum in the dependence of the stellar mass on the
central density $M_\textrm{s} (\rho_c)$ for $T=0$.
The mass of such a spherical star is $M=1,0042M_\odot$.

To define the initial model the density (and hence the mass) of the star was increased by 20\% at
every point.
The temperature in the star was defined by the relation: $T=\delta\rho^{2/3}$, where
$\delta =1(\textrm{K}\cdot\textrm{cm}^2\cdot\textrm{g}^{-2/3})$.
At the initial time moment $t=0$, the rigid-body rotation with the angular velocity
$\omega=2.519\>\textrm{c}^{-1}$ (rotational period is  $\tau=2.496\>\textrm{c}$) was accepted.
At $t=0$ we suppose also that  $v_r=v_z=0$.
The initial rotational energy is  0.571\% of the gravitational energy,
and the initial internal energy including
the energy of degeneracy is 72.7\% of the gravitational energy.
The defined initial model is unstable against collapse, and immediately after
the beginning of the calculations it starts to contract.

The initial magnetic field (the initial magnetic field which was "turned on" after
the collapse stage) was defined exactly in the same way as in \citet{abkm}.
We define the toroidal current $j_\varphi$ in the following form:
\begin{equation}
j_\varphi=\left\{
\begin{array}{l}
j_\varphi^u \> {\rm for} \> z\geq 0,\> r^2+z^2\leq 0.025^2, \\
j_\varphi^d \> {\rm for} \> z\leq 0,\> r^2+z^2\leq 0.025^2,\\
0 \> \> \> \>{\rm for} \> {\phantom {z\leq 0,}}  \>r^2+z^2\geq 0.025^2,
\end{array} \right. \label{inicur}\\
\end{equation}
where
\begin{eqnarray*}
j_\varphi^u  = A_j\Big[{\rm sin}\left(\pi {r\over 0.025} -
              {\pi \over 2}\right)+1\Big]
          \Big[{\rm sin}\left(\pi {z\over 0.025} -
              {\pi \over 2}\right)+1\Big]\nonumber\\
       {}  \times\Big[1-\left({r\over 0.025}\right)^2
            -\left({z\over 0.025}\right)^2\Big],\nonumber\\
j_\varphi^d = -j_\varphi^u.{}\nonumber
\end{eqnarray*}
$A_j$ is a coefficient used for adjusting the values of the initial toroidal current
and, hence, the magnetic field.
After the collapse of the core and obtaining the differentially
rotating stationary configuration without a magnetic field.
we calculated the initial poloidal magnetic field $H_{r0},H_{z0}$
using the Bio-Savara law. The calculated magnetic field is divergence-free but not force-free
and not balanced with
other forces. To make it balanced we use the following method (\citealt{abkm}):
we "turn on" the poloidal magnetic field
$H_{r0},\>H_{z0}$, but "switch off" the equation for the
evolution of the toroidal component $H_\varphi$ in
(\ref{magmain}). In fact this means that we define $H_\varphi
\equiv 0,\> {{\rm d}H_\varphi \over {\rm d} t} \equiv 0$.  From
the physical point of view it means that we allow magnetic field
lines to slip through the matter of the star in the toroidal
direction $\varphi$.
After "turning on" such a field, we let the cloud
come to a steady state, where magnetic forces connected
with the purely poloidal field are balanced by other forces.
The calculated balanced configuration has a magnetic field of
quadrupole-like symmetry. The relation of the magnetic energy of the star to its gravitational
energy at the moment of the formation of the balanced poloidal magnetic field is $10^{-6}$.

\section{Numerical method}
The implicit operator-difference completely
conservative scheme on the triangular grid of variable structure was used for
the numerical modeling of  the problem of the magnetorotational
supernova explosion.
The  scheme was suggested and investigated  by \citet{arkoche} and
in earlier papers by these authors.

The solution of the problem is a sequence of  time steps.
Calculation of every time step can be divided into two parts.

The first part is the calculation of the values of the functions
on the next time level using the implicit completely conservative
operator-difference scheme on the triangular grid
in Lagrangian variables \citet{arko}, \citet{arkoche}. The
coordinates of grid knots are changing at this stage.

The second part is an analysis of the quality of the grid, its
improvement and adaptation (grid reconstruction).
The improvement of the quality of the grid is necessary because
of the appearance of "poor" cells, i.e. triangles which
strongly deviate from equilateral triangles.
The dynamical adaptation of the grid allows us to concentrate the
grid in the regions of the computational domain where spatial
resolution needs to be increased and rarefy the grid in the
regions where the flow is smooth. It allows us to reduce significantly
the dimensionality of the grid and hence strongly reduce the computation time.

The grid reconstruction procedure itself consists of the following stages.
The first stage is a local correction of the structure of the grid. The second
stage is a calculation of the values of the functions defined in cells
in the regions of the corrected structure.

The local correction of the structure of the grid can be done using the following three
local operations (see details in \citet{abkkm}):
\begin{enumerate}
  \item replacement of the diagonal of the quadrangle formed by two triangles by  another diagonal;
  \item joining up of two neighboring  grid knots;
  \item adding of the knot at the middle of the cell side which connects two knots.
\end{enumerate}

The improvement of the grid structure is made using the first two operations. The grid adaptation
is made by application of the local operations (II) and (III). The plots which clearly explain
these operations are given in \citet{abkkm}.

The values of functions defined in cells and knots involved in grid structure modification
are calculated at the second stage of the grid reconstruction.

The application of simple interpolation for the calculations of  new
values of the grid function leads to the violation of the conservation laws and
adds significant errors in the regions with high gradients (for example at  shock waves).
The goal of the calculation of the new values of functions is to minimize numerical errors
introduced by this procedure.
To achieve this goal not only the error in the values of the functions
but also on their gradients have to be minimized. It is
important to fulfill conservation laws (mass, momentum, energy, magnetic flux)
in the vicinity of the local grid reconstruction. The method for the calculation
of these new values of grid functions is based on a minimization
of the functionals containing the values of the functions, its gradients, and
grid analogs of the conservation laws (\citealt{arko}).

For the solution we have
used the method of the conditional minimization of the functionals guaranteeing
exact fulfilment of the conservation laws.

It is important to fulfill conservation laws for the solution of the collapse  and
magnetorotational supernova explosion problems, because a large number of the time steps
need to be made. In such a situation even slight violation of the conservation laws at a time step
could lead to a significant growth in  the errors and hence to a qualitative distortion of the results.

The method of calculating hydrodynamical values $\rho$ and $P$ during grid reconstruction
is taken from \citet{arko}. To obtain a better precision some
changes have been made in the calculation of the magnetic field
components in comparison with our calculations of the collapse of
magnetized cloud (\citet{abkm}). Earlier, for the calculation of the new
values of the magnetic field components we conserved the sum of the products
of the magnetic field components values  and the volumes of the cells in the
vicinity of the reconstructed part of the grid. In this paper we conserve
the poloidal magnetic energy  and the toroidal magnetic flux
in the vicinity of the reconstructed part of the grid.

The grid reconstruction procedure allows us not only to "correct"
the Lagrangian grid, but also to dynamically adapt it
using different criteria for the grid in different
parts of the computational domain.
The grid can be refined in the regions where it is
necessary and therefore we can increase the accuracy of the calculations.
It is possible to rarefy the grid in those parts of the computational
domain where the flow is "smooth". The procedure of the
rarefying of the grid allows us to significantly reduce the dimension
of the grid while preserving the same accuracy for the numerical solution.

Geometrical criteria can be used for the grid adaptation (i.e.
restrictions on the length of the cell size, which are defined by
the cell coordinates only), but such adaptation criteria are suitable
for the flows of simple or easily predictable structures only.

For grid adaptation in the case of complicated flows or flows with
unknown structures it is better to apply dynamical criteria which
are defined by the solution behavior. The dynamical criteria for
the grid adaptation applied here was suggested by \cite{abkkm}.

The characteristic length of the cell side $l_k$ was chosen as a local criterion
for the grid reconstruction. As an example, consider the criterion where $l_k$
is a function of $\rho$ and grad$\rho$.
Let's introduce the function:
\begin{equation}\label{critad}
  f(\rho, {\rm grad} \rho)=\frac{\alpha}{(\rho+\varepsilon)^{\gamma_1}}+
  \frac{\beta}{(|{\rm grad}\rho|+\varepsilon)^{\gamma_2}},
\end{equation}
where $0<\varepsilon<<1, \> \alpha\geq 0, \> \beta \geq 0,\> \alpha+\beta=1$.
$\gamma_1,\> \gamma_2$ are power indexes, and ${\rm grad}\rho$ is the grid analog of the density gradient.
In limited cases:

$\alpha=1,\> \beta=0, \> f$ depends on the density only;

$\alpha=0,\> \beta=1, \> f$ depends on the gradient of the density only.

Let $N$ be the total number of the grid cells.
The characteristic length of the side $l_k$ of the cell with
the number $k$ will be calculated as a function
of the density, the gradient of the density, and the coordinates
$r,\> z$ (implicitly) by the following formula
\begin{equation}\label{dlisto}
  l_k=2\sqrt{\frac{s_k}{3}}, \> \>
  s_k=\frac{f(\rho_k,\>{\rm grad}\rho_k)}
  {\sum\limits_{n=1}^{N} f(\rho_n, \>{\rm grad} \rho_n)} S.
\end{equation}
Here, $S$ is a square of the computational domain, which consists of the
triangular cells. The summation in the denominator is made for all grid cells.
Note that $S=\sum \limits_{k=1}^{N} s_k$, where
$s_k$ is equal to the square of the equilateral triangle with the length of the side $l_k$.
In our calculations the function $f$ in formula (\ref{critad}) was used in the following form
$f(\rho_k)=1/(\rho_k+\varepsilon)^{0.5}$.

The criterion described was applied in the following way.
In the case where the length of the side of the cell $i$ is larger
than $2l_k$, a new knot is added to the middle of this
side of the cell. In the case where the length of the side of the cell $i$ is
less than $0.7l_k$, the operation of the joining up of these knots is applied.
The application of the dynamical adaptation criterion described
allowed us not only to adapt the grid to the specialities of the
solution but also to provide an acceptable accuracy of the
calculations with a small fluctuation of the total number of
 grid knots and cells. In the calculations described the
total number of knots (5000) and cells (10000) was violated not more
than 5\%.

At the moment of the maximum compression (for the collapse
problem) the minimal size of the cell side is so small that
application of the uniform grid with the same spatial resolution
would require a grid with the dimension $\sim 1000\>  \times\> 1000 $ (!) cells.

The calculation of the gravitational potential in this paper as in
 \citet{abkkm2004} in the frames of the applied numerical
method is made on the base of the finite element method of higher
order (\citet{zemo}). This procedure allowed us to increase the accuracy
of the calculations and to eliminate the loss of approximation near the $z$ axis.

The linear artificial viscosity (e.g., \citealt{sampop}) was introduced
into the numerical scheme for the shock capturing.

\section{Results}
The simulation of the magnetorotational supernova is divided into
two separate parts. The first part is the core collapse simulation,
and the second is the "switching on" of the initial poloidal
magnetic field following
amplification of its toroidal
component $H_\varphi$, which is finished by the magnetorotational explosion.
The duration of the second phase is defined by
the value of the initial magnetic field. The weaker the initial poloidal
magnetic field, the longer the stage of the amplification
of the toroidal component until the supernova explosion. The results of
1D simulations of the magnetorotational mechanism made by
\citet{bkpopsam} and \citet{abkp} show that the time from the beginning
of the magnetic field evolution to the explosion is proportional to
$1/\sqrt{\frac {E_{mag0}}{E_{grav0}}}$,
where $E_{mag0}$ and $E_{grav0}$ are initial magnetic and
gravitational energies of the star, respectively.

\subsection{Core collapse simulation}
The core collapse simulation  was
described in detail in  \citet{abkkm2004}. In the present paper we
describe briefly the obtained results.
The ratios between the initial rotational and gravitational energies and between
the internal and gravitational energies of the star are the following:
$$
\frac{E_{rot}}{E_{grav}}=0.0057, \> \frac{E_{int}}{E_{grav}}=0.727.
$$
Soon after the beginning of the contraction at $t=0.1377$s at the
distance ~$6\cdot 10^5$cm the bounce shock wave appears.
Behind the shock front the temperature of the matter rises sharply and
neutrino losses are "switching on". In Fig. \ref{neutlos} the time
evolution of the neutrino losses is presented.
\begin{figure}
\centerline{\includegraphics{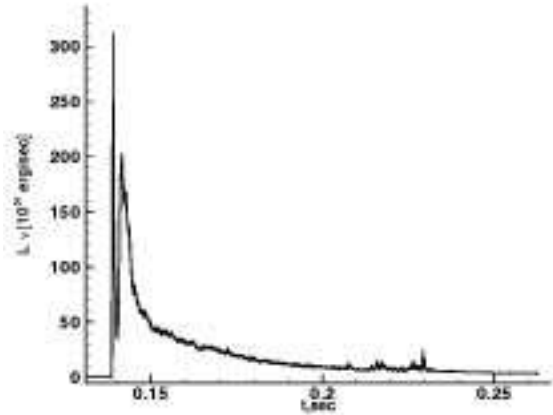}} \caption{Time evolution
of the neutrino luminosity $\int\limits_0^{M_{core}} f(\rho,T) dm$
during collapse.}
\label{neutlos}
\end{figure}
The matter of the star from the outside of the shock wave continues to
collapse towards the center of the star.

At $t=0.1424$s the density in the center of the star reaches its
maximum value $\rho_{c,max}=5.655\cdot10^{14}$g/cm$^3$. The matter
of the envelope which has passed through the bounce shock wave forms the
core (proto-neutron star). Behind the shock front the intensive mixing of the
matter takes place.

The shock wave moves through the envelope and at $t=0.2565$s reaches the outer
boundary of our computational domain near rotational axis $z$. The shock leads
to the ejection of  $0.041\%$ of the core mass and $0.0012\%$ ($2.960\cdot10^{48}$erg)
of the gravitational energy of the star. The particles of the matter are treated
as "ejected" when their kinetic energy becomes larger than their gravitational
energy and the velocity vector is directed from the star center ($r=0,\>z=0$).
The amounts of the ejected mass and energy are too small to explain the core collapse supernova
explosion. It should be noted that in  \citet{jankaplewa},
where the neutrino losses were calculated using the solution of the Boltzmann equation, the
shock wave does not lead to the ejection of the matter.
At the final stage of the core collapse ($t=0.261$s.) we obtain a
differentially rotating configuration. The central proto-neutron star with a
radius $\sim12.8$km rotates almost rigidly with the rotational period 0.00152s.
The angular velocity rapidly decreases with the increase in the distance from the star center.
The particles of the matter situated at the outer boundary in the equatorial
plain rotate with the period $\sim 35$s (Fig. \ref{period}).
\begin{figure}
\centerline{\includegraphics{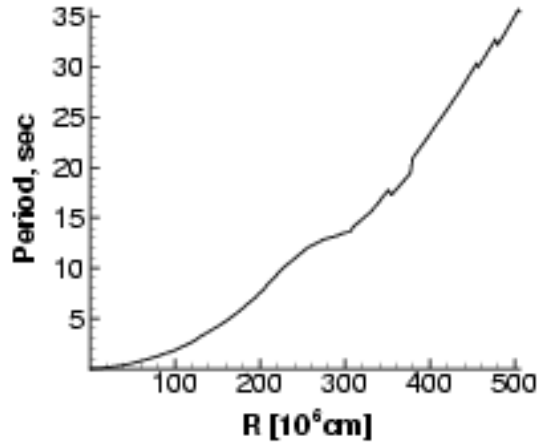}}
 \caption{Dependence of the rotational period (in s.) from $r$
 coordinate in the equatorial plane ($z=0$) after the collapse for $t=0.261$s.}
  \label{period}
\end{figure}

\subsection{Magnetorotational explosion}
The results of the collapse simulations show that the amounts
of the  mass and energy ejected by the bounce shock wave are
too small to explain the supernova explosion.

The small ejecting part of the envelope prevents us from
continuing the calculations because of significant increase
in the square of the computational domain. To overcome this problem the
outer part of the envelope was stopped by artificially reducing
its velocity (kinetic energy). This artificial method does not influence
the central parts of the collapsed core.

After "stopping" the ejecting part of the envelope we run our code
without the magnetic field to be sure that the
configuration formed is not changing in time significantly.
We reach the stage where the poloidal kinetic energy of the star is less than $0.00001\%$ of the
gravitational energy of the star and  remains below that value
for 1000 time steps. The shape of the envelope of the star does not
change significantly during this test.

At that stage we  "switch on" the initial poloidal magnetic field, defined by the
current (\ref{inicur}). The energy of this magnetic field was taken to be
equal to $\sim10^{-6}$ of the gravitational energy of the star
at the moment of including the magnetic field in our simulations.
The poloidal magnetic field, defined by the current (\ref{inicur}),
is divergence-free, but is not force-free, and "switching on"
that field can lead to the artificial violation of the equilibrium of the star.
To reach a steady state of differentially rotating configuration with the balanced
poloidal magnetic field we exactly follow the procedure from \citet{abkm}.

The balanced configuration calculated has the magnetic field of
quadrupole-like symmetry. After the formation of the
balanced poloidal magnetized configuration we "switch on" the
equation for the evolution of the toroidal component $H_\varphi$.

We calculate the evolution of the magnetic field after the collapse stage, because
at the developed stage the collapse is rather short in time and newly forming proto-neutron star
rotates not as differentially as at the end of the collapse. During the collapse
the forming proto-neutron star makes only a few revolutions and the magnetic field,
which is initially weak, does not have a significant influence on the flow in the star.

At the moment of "switching on" the toroidal magnetic field we start a
counting the time anew.

At the beginning of the simulations the toroidal component of the
magnetic field grows linearly with the time  at the periphery of the
proto-neutron star. The energy of the toroidal magnetic field grows
as a quadratic function (Fig. \ref{magenrotpol}). At the developed stage
of the $H_\varphi$ evolution ($t=0.04s$) the poloidal magnetic energy begins to grow
 much faster due to developing magnetorotational
instability (\citealt{akiyama}) leading also to a rapid growth of the poloidal components.
\begin{figure}
\centerline{\includegraphics{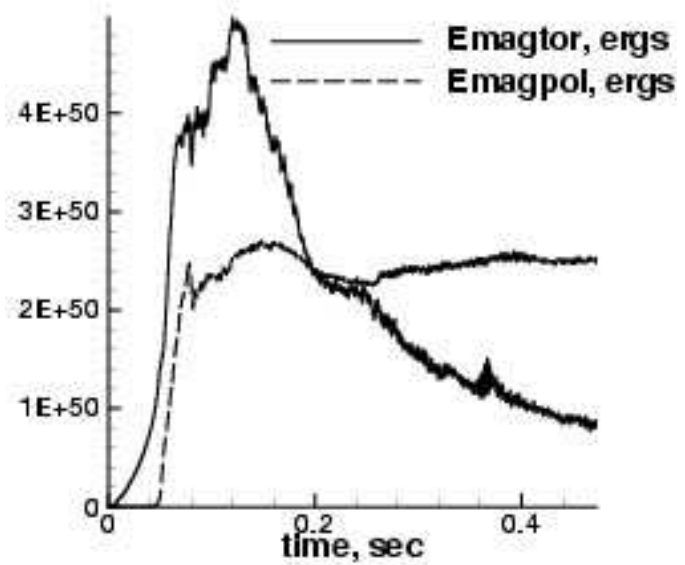}}
 \caption{Time evolution of the toroidal and poloidal
 parts of the magnetic energy during magnetorotational explosion.}
  \label{magenrotpol}
\end{figure}
\begin{figure}
\centerline{\includegraphics{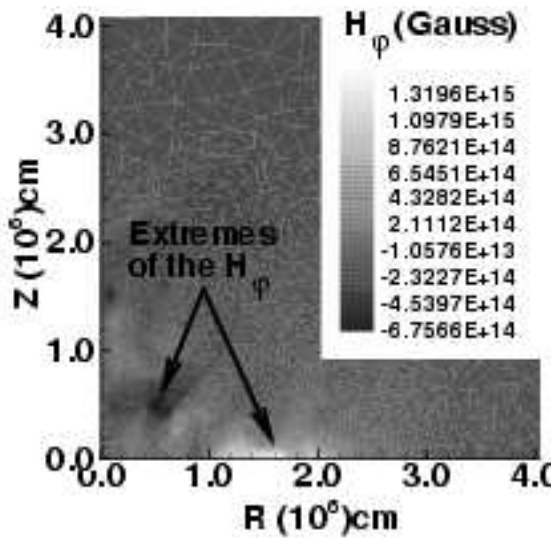}}
 \caption{The toroidal field $H_\varphi$ distribution
 after $0.012$s of the beginning of the evolution of the
 $H_\varphi$ in the central part of the star.}
  \label{hf0bw}
\end{figure}
Due to the quadrupole-like type of the symmetry of the initial
magnetic field the generated toroidal field $H_\varphi$ has two
extremes. The first is at the equatorial plane  and the second is out of the equatorial plane
at the periphery of the proto-neutron star closer to the axis of rotation $z$.
The two extremes have different signs of $H_\varphi$. In
Fig. \ref{hf0bw} the distribution of the $H_\varphi$ is plotted at $0.012$s.
 These extremes approximately
correspond to the extremes of the  term
$r{\bf H} \cdot {\textrm grad} (V_\varphi /r)$ (\citealt{abkm})
in the equation for the evolution of the $H_\varphi$, because the star
is in a steady state condition, and
only this term in this case determines the evolution of $H_\varphi$.

The maximal value of the $H_\varphi$ reached during the amplification
of the toroidal field phase is $\sim  2.5\cdot 10^{16}$G.
This maximum is situated at the equatorial plane at a distance
$1.7\cdot 10^{6}$ cm from the center of the star, and it is reached at
$t=0.058$s. The toroidal part
of the magnetic energy decreases with time after reaching its maximal value of
$4.8 \cdot 10^{50}$ergs at $t=0.12$s.
The poloidal magnetic energy at the developed explosion stage is
$\approx 2.5 \cdot 10^{50}$erg and keeps this value until the end of our
simulations (Fig. \ref{hf0bw}).

The magnetic pressure is highest in the regions where the $H_\varphi$
reaches  its extremal values. At these regions the $|H\varphi|$ is approximately
100 times higher than the absolute value of the poloidal field
($\sqrt{H^2_r+H^2_z}$).

The angular momentum is subtracted by the
magnetic torque from the proto-neutron star. The magnetic field works as a
"transmission belt" for the angular momentum. The envelope of the star
starts to blow up slowly. The contraction wave appears at the periphery of the
proto-neutron star near the extremes of the $H_\varphi$.
This contraction wave moves along a steeply decreasing density profile.
The amplitude of this wave rapidly grows, and after a short time it transforms
into the fast MHD shock wave. The growing toroidal magnetic field due to the
differential rotation works as a piston for the originated MHD shock.
Time evolution of the velocity field $v_r,\>v_z$, specific angular
momentum $v_\varphi r$, and temperature $T$ for the time moments
$t=0.07,0.20,0.30s$ is given in  Fig. \ref{angmom}, and Fig. \ref{veltemp}.
\begin{figure}
  \centerline{\includegraphics{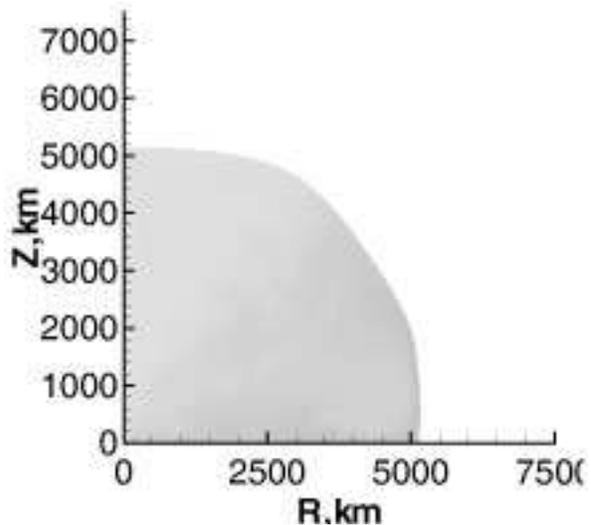}}
    \vspace*{-15pt}
  \centerline{\includegraphics{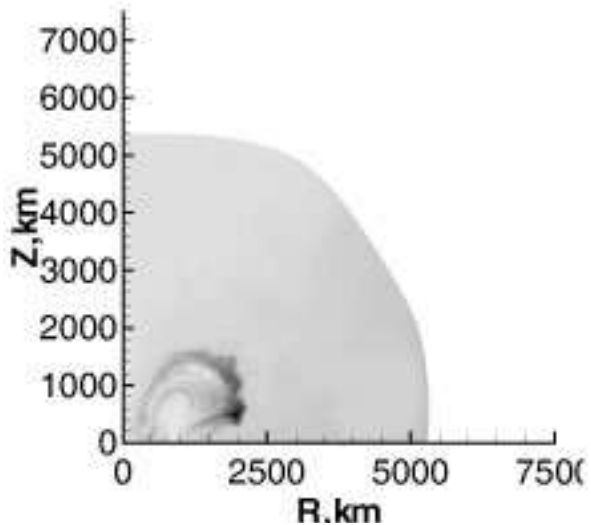}}
    \vspace*{-15pt}
  \centerline{\includegraphics{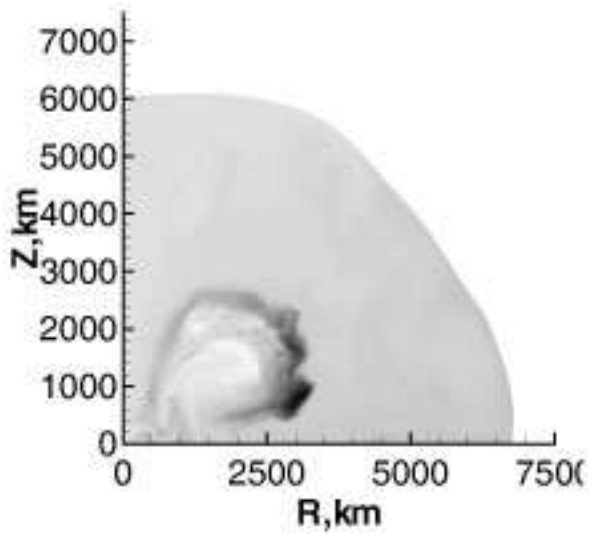}}
  \caption{Time evolution of the specific angular momentum $v_\varphi r$
  for the time moments $t=0.07s,\>0.20s,\>0.30s$. The darker parts of the plots correspond to the
  higher specific angular momentum.}
  \label{angmom}
\end{figure}
\begin{figure*}
  \centerline{\includegraphics{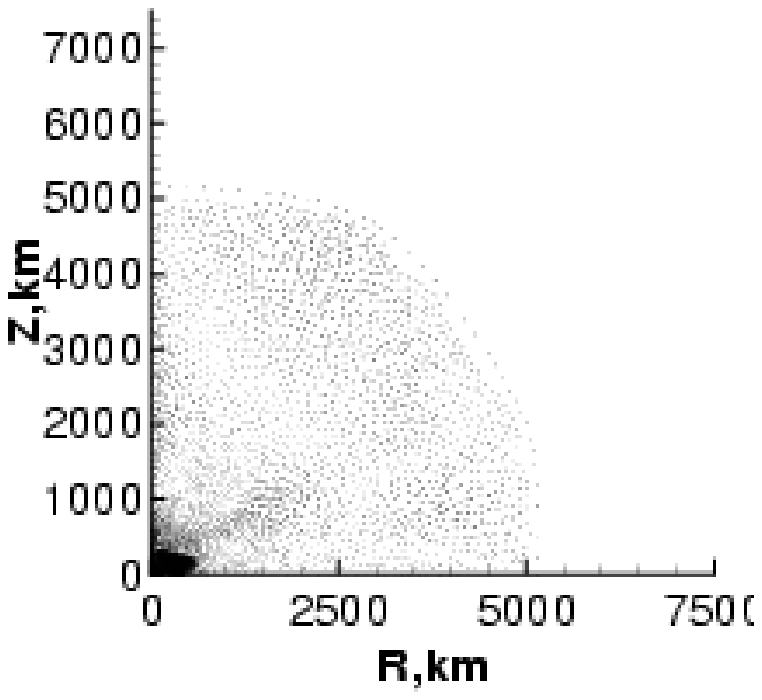}\includegraphics{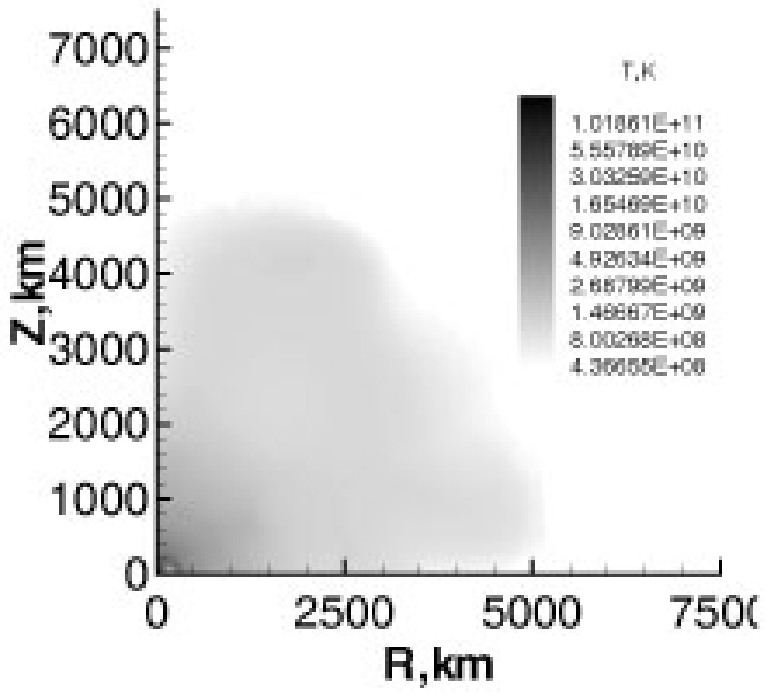}}
    \vspace*{-10pt}
  \centerline{\includegraphics{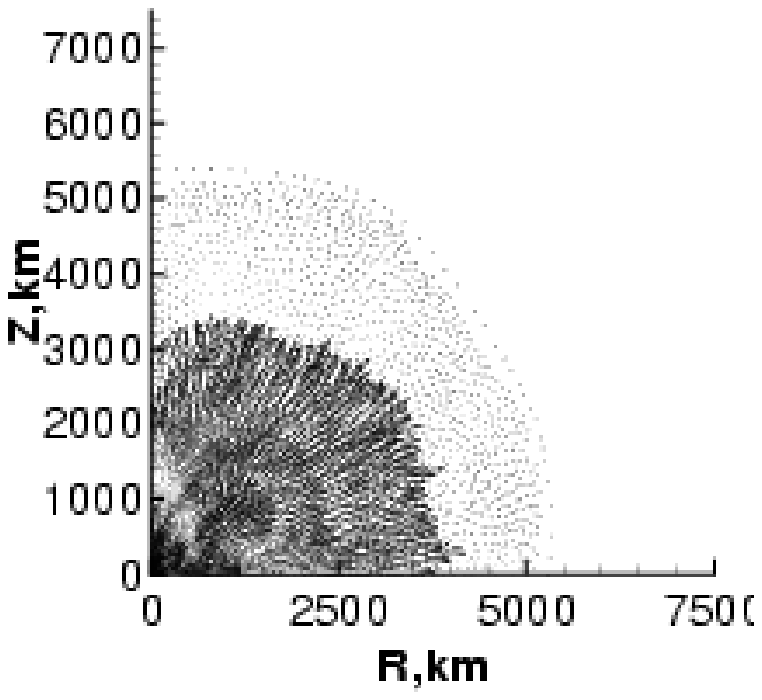}\includegraphics{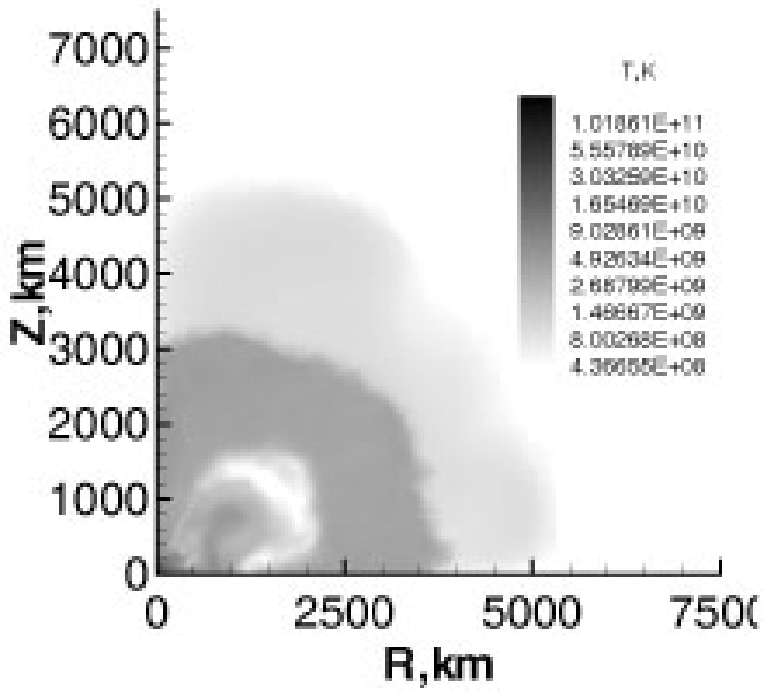}}
    \vspace*{-10pt}
  \centerline{\includegraphics{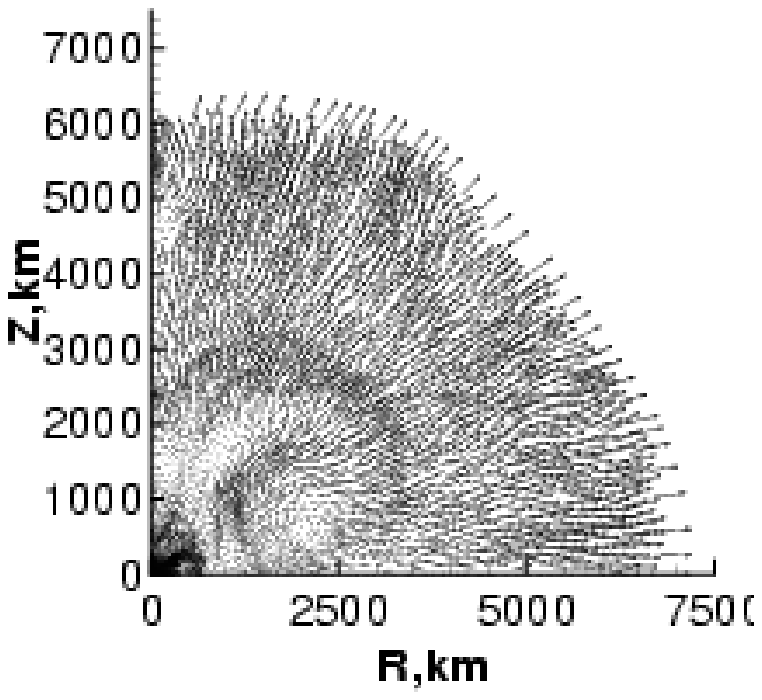}\includegraphics{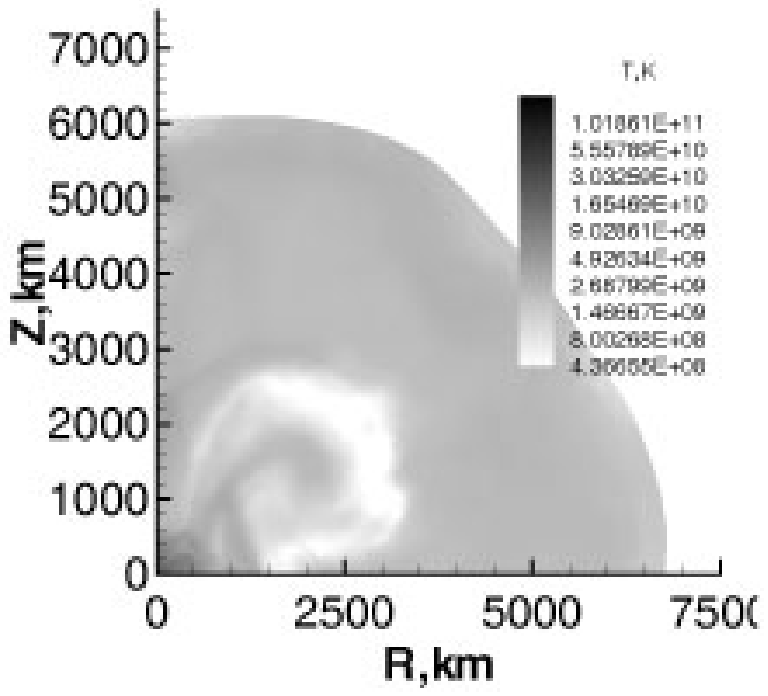}}
  \caption{Time evolution of the velocity field (left column) and temperature
  (right column) for the time moments $t=0.07s,\>0.20s,\>0.30s$.}
  \label{veltemp}
\end{figure*}

The flow behind the initial MHD shock is very inhomogeneous. Possible
reasons of this inhomogeneity are:
\begin{itemize}
  \item the aftershock behavior of the gas flow in the
presence of the gravitational field,
  \item the magnetorotational instability appearing at the periphery
of the proto-neutron star.
\end{itemize}

  The oscillating structure of the flow behind the shock wave in
the gravitational field (without magnetic fields) was
investigated analytically in a 1D case in
the acoustical approximation by \citet{lamb} and numerically for
1D gravitational gas dynamics by \citet{kospop}. Similar oscillating structure
was found in our 2D simulations of the collapse of the rapidly rotating
cold protostellar cloud (\citealt{abkkm}). The main reason for this effect
is a change of dispersion properties of the matter in the presence of a
gravitational field.

During the evolution of the magnetorotational explosion the proto-neutron star
loses angular momentum, contracts  and slightly increases its rotation.
This leads
to the further
amplification of the toroidal magnetic field in this region. In other words
it means that the "piston" continues to push the matter from the central object to
the envelope and to support the MHD shock. The continuous support of the supernova shock
in the magnetorotational mechanism is the main qualitative difference from the prompt
shock and neutrino driven supernova mechanisms.

Due to the quadrupole-like type of the symmetry of the initial magnetic field the
MHD shock is stronger, and it moves faster near the equatorial plane $z=0$.
The matter of the envelope of the star is ejected preferably near the equatorial plane.

The formation of the MHD shock and its propagation leads to the secondary neutrino losses
jump (Fig. \ref{neutlum}), but now it is much weaker than it was at the collapse
stage.
\begin{figure}
\centerline{\includegraphics{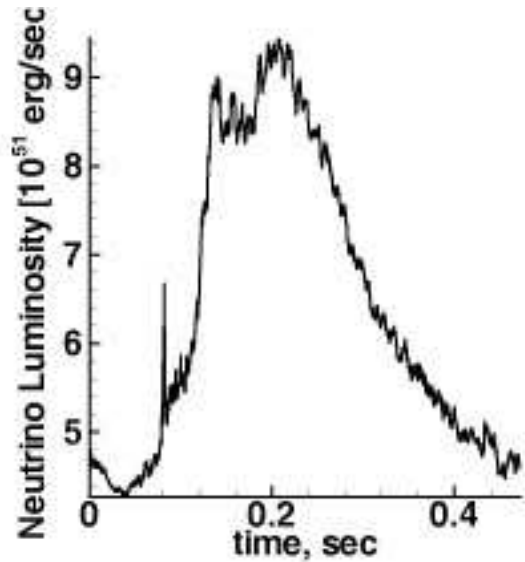}}
 \caption{Time dependence of the neutrino luminosity $\int\limits_0^{M_{core}} f(\rho,T) dm$
 during the magnetorotational explosion.}
  \label{neutlum}
\end{figure}

The integral neutrino losses as a function of time (during magnetorotational stage) is given in the Fig. \ref{neutlosm}.
\begin{figure}
\centerline{\includegraphics{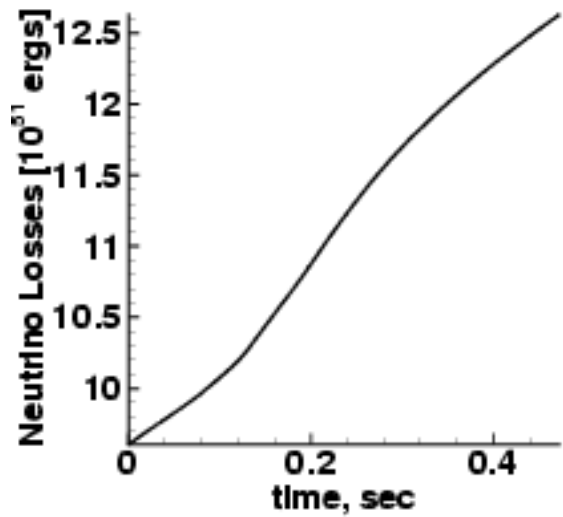}}
 \caption{The integral neutrino losses as a function of time
          during the magnetorotational explosion.}
  \label{neutlosm}
\end{figure}

Time dependence of the ejected mass of the star is given in the Fig. \ref{ejmas}.
\begin{figure}
\centerline{\includegraphics{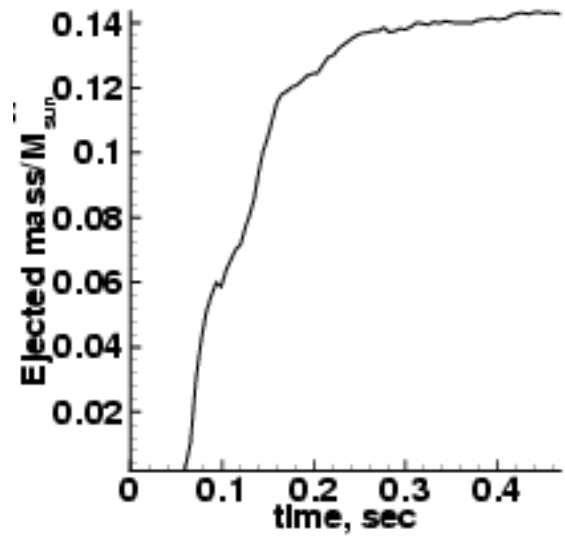}}
 \caption{Time dependence of the ejected mass in relation to $M_\odot$  during the magnetorotational explosion.}
  \label{ejmas}
\end{figure}
In Fig. \ref{ejenergy} the time dependence of the ejected energy is represented. The
particle is considered  as УejectedФ  if its kinetic energy is greater than its potential energy
and its velocity vector is directed from the center.
\begin{figure}
\centerline{\includegraphics{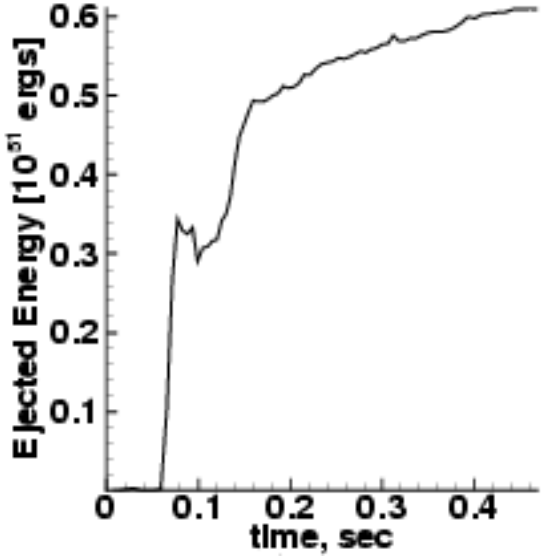}}
 \caption{Time dependence of the ejected energy during magnetorotational explosion.}
  \label{ejenergy}
\end{figure}
The results of our simulations show that during magnetorotational explosion
$\sim 0.14\>M_\odot$ of the mass  and  $\sim 0.6\cdot 10^{51}$ergs are ejected.

The MHD shock which produces the supernova is a fast MHD shock (as in \citealt{abkm}),
because its velocity is larger than fast magnetic sound speed in the upstream
flow, while the velocity of the slow MHD shock in the downstream flow is between
Alfvenic and slow magnetic sound speeds.

The magnetic field  transforms part ($\sim 10\%$) of the rotational
energy of the star to the radial kinetic energy (explosion energy).
The time dependence of the rotational energy and the poloidal part of the
kinetic energy are given in Fig. \ref{rotkinpolen}.
\begin{figure}
\centerline{\includegraphics{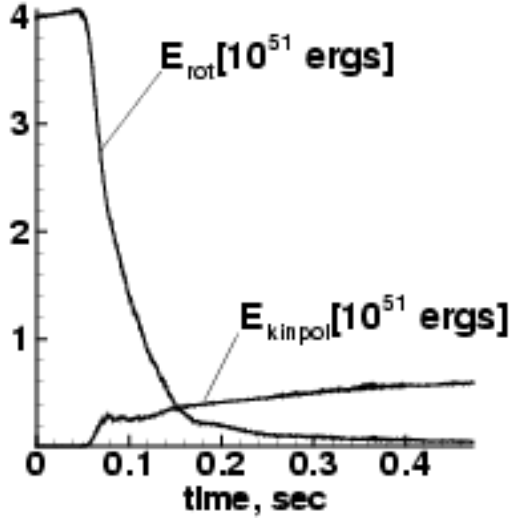}}
 \caption{Time dependence of the rotational energy and poloidal kinetic energy
 of the star during magnetorotational explosion.}
  \label{rotkinpolen}
\end{figure}
\begin{figure}
\centerline{\includegraphics{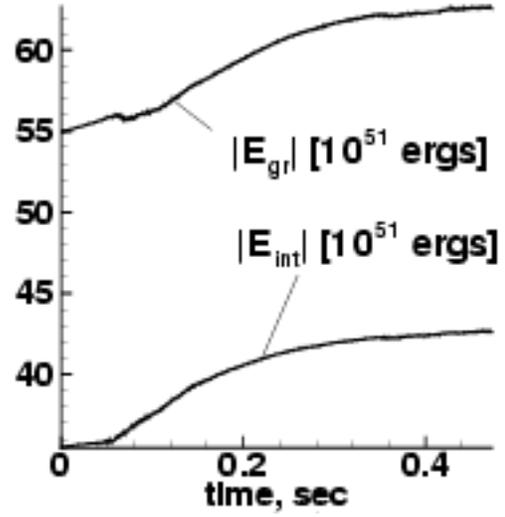}}
 \caption{Time dependence of the gravitational and internal energies
 of the star during magnetorotational explosion.}
  \label{grinten}
\end{figure}

During the magnetorotational explosion the star loses a significant part of its
rotational energy. The rotational energy of the star is transformed
not only into the explosion (kinetic energy of radial motion) but is lost in the form of
neutrino emission and partially changes the total energy of the star (Fig. \ref{grinten}).
The core rotates more slowly now, and it leads to a deeper contraction and
some heating of the proto-neutron star.

We stop calculations at $t \approx 9.45 s$. At this stage the
proto-neutron star rotates with the period $\sim 0.006$s. The
absolute value of the poloidal magnetic field at the periphery of the proto-neutron
star (at the equatorial plane, $\sim 10$km from the center of the star) is
$\sqrt{H_r^2+H_z^2}\approx 2\cdot 10^{14}$G.

\section{Appearance of the magnetorotational instability in 2-D picture}
Magnetorotational instability (MRI) in the magnetized star with
differential rotation was analyzed by \citet{spruit}, who
indicated to dynamo action, accompanies a development of such
instability. In our axially symmetric 2D simulations dynamo action
is prohibited (\citealt{cowling}), nevertheless the development of
the MRI takes place. It was investigated in 2D numerically for the case
of accretion discs by \citet{hawbalb} and
\citet{fromang}. The possibility of the appearance of such
instability was also mentioned by \citet{colgate} in application
to the SN1987a.  The qualitative picture of the MRI in 2D is
the following.  At the first stages the differential rotation
leads to a linear growth of the toroidal field described
qualitatively as
\begin{equation}\label{mri1}
\frac{{\mathrm d} H_\varphi}{{\mathrm d} t}=
H_r\left( r\frac{{\mathrm d} \Omega}{{\mathrm d}r}\right).
\end{equation}
The right side is constant at the initial stage of the process.
When toroidal field reaches its critical value $H_{\varphi *}$,
the MRI instability starts to develop. As follows from our
calculations the critical value corresponds to the relation
between toroidal magnetic and internal energy densities as
\begin{equation}
  \frac{H^{* 2}_{\varphi }}{8\pi}=\frac{1}{10}E\rho.
\end{equation}
Appearance of MRI is characterized by formation of multiple {\it
poloidal} differentially rotating vortexes, which twist the
initial poloidal field leading to its amplification according to
\begin{equation}\label{mri2}
  \frac{{\mathrm d}H_r}{{\mathrm d} t}=
  H_{r 0}\left(\frac{{\mathrm d} \omega_v}{{\mathrm d} l} l \right),
\end{equation}
where $l$ is the coordinate, directed along the vortex radius,
$\omega_v$ is the angular velocity of the poloidal vortex.
Qualitatively the poloidal field amplification due to the vortexes
induced by MRI is shown in the Fig. \ref{mrifig}.
\begin{figure}
\centerline{\includegraphics[width=0.4\textwidth]{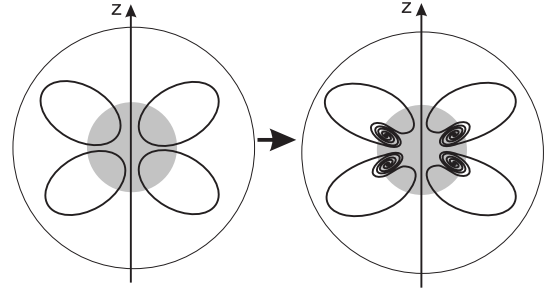}}
\hfill
 \caption{Qualitative picture of the development of the magnetorotational instability.}
  \label{mrifig}
\end{figure}
The enhanced poloidal field immediately starts to take part in the
toroidal field amplification according to (\ref{mri1}). With
further growing of $H_\varphi$ the poloidal vortex speed
increases. Our calculations give the values of $\omega_v=0.0132\>
{\mathrm s^{-1}}$ at $|H_\varphi|=2.46 \cdot 10^{15}{\mathrm G}$
corresponding to $t=0.041 \>{\mathrm s}$ and $\omega_v=0.052\>
{\mathrm s^{-1}}$ at $|H_\varphi|=4.25\cdot 10^{15}{\mathrm G} $
corresponding to $t=0.052 \>{\mathrm s^{-1}}$ for the same
Lagrangian particle. In general we may approximate the value in
brackets (\ref{mri2}) by linear function on the value
$(H_\varphi-H^*_\varphi )$ as
\begin{equation}\label{mri3}
  \left(\frac{{\mathrm d} \omega_v}{{\mathrm d} l} l \right)= \alpha (H_\varphi-H^*_\varphi ).
\end{equation}
$H_r$(or $H_{poloidal}$ in general case) in (\ref{mri1}) is not
constant anymore after onset of MRI, but described by (\ref{mri2})
and (\ref{mri3}). Assuming for the simplicity that $\left( r
\frac{{\mathrm{d}\Omega}}{{\mathrm d}r}\right)=A$ is constant
during the first stages of MRI, and taking $H^*_\varphi$ as constant
we come to the following equation:
\begin{equation}\label{mri4}
  \frac{{\mathrm d}^2}{{\mathrm d} t^2}\left( H_\varphi - H^*_{\varphi }\right)=
  A H_{p 0} \alpha (H_\varphi - H^*_{\varphi }),
\end{equation}
giving the exponential growth of magnetic field components

\begin{eqnarray*}
  H_\varphi=H^*_{\varphi } + H_{p0} e^{\sqrt{A\alpha H_{p0}}(t-t^*)}, \\
  H_{r}=H_{r0}+\frac{H^{3/2}_{p0} \alpha^{1/2}}{\sqrt{A}}\left( e^{\sqrt{A\alpha H_{p0}}(t-t^*)} -1 \right),
\end{eqnarray*}

\noindent taking $H_{p0}$ as seed field for development of MRI.

The above toy model shows the way of development of MRI instability
in 2D case. In this case there is no direct influence of toroidal
field onto a poloidal one, therefore there is no dynamo action.
Nevertheless, both components grow exponentially, chaotically for the
poloidal field and both chaotically and regularly for the toroidal one.
Magnetorotational explosion is produced by both, regular and chaotic
magnetic fields. After rapid neutrino cooling and damping of the hydrodynamic
motion, the chaotic component remains
frozen into the matter due to high electric conductivity, and chaotic magnetic field
could be the source of magnetar activity in soft gamma repeaters \citet{belobor}.

\section{Discussion}
Our calculations have shown that the magnetorotational mechanism produces enough energy
($0.6 \cdot 10^{51}$ergs)
to explain the core-collapse supernova of types II and Ib. The Ic type
is probably more energetic and could be connected with the collapse  of much more
massive cores of a few tens of solar masses. The value of the energy production in this particular variant
may be considered as a basic one, which may be several times larger or smaller, depending on the mass of the
collapsing core, magnetic field intensity and configuration, and rotational energy of the presupernova.

The magnetorotational explosion of a supernova star is divided into three stages:
\begin{enumerate}
  \item linear growth of the toroidal magnetic field due to the twisting of magnetic field lines,
  \item exponential growth of the toroidal and poloidal magnetic fields due to the development of
magnetohydrodynamic instabilities (\citealt{dungey}, \citealt{tayler}, \citealt{spruit}),
  \item formation of the MHD shock wave and magnetorotational explosion.
\end{enumerate}

The resulting newborn neutron star is characterized by relatively slow rotation: 6 ms in
comparison to sub-ms values of the critical rotation. The resulting period could grow
with increasing efficiency of the explosion.

The toroidal magnetic energy of the young neutron star decreases more rapidly than the poloidal
component which tends to the constant value (see Fig. \ref{magenrotpol}). Possibility
of the dynamo action in a similar situation, leading to an increase in the large scale poloidal as
well as toroidal components, was considered by \citet{spruit}. In our calculations
the dynamo action is prohibited
due to 2D geometry (\citealt{cowling}).

Different variants of magnetorotational instability in astrophysics were investigated in
connection with jet formation (\citealt{lovelace}, \citealt{meiernakamura}),
and generation of turbulence  in the accretion disks (\citealt{balbus}).
In the last case the instability of the uniform field,
threading the differentially rotating disk is considered,
which was investigated earlier in plasma physics (\citealt{velikhov}, \citealt{chandr}).

The large values of the remaining magnetic field in our case are
related to the chaotic, nonregular components,
which have zeroaverage magnetic flux and could disappear by field
annihilation  even at large conductivity.
The existence of a large chaotic component of magnetic field on
a neutron star may last long due to very high conductivity. It may be
related the magnetars model or soft gamma repeaters, in which the
radiated energy comes from the magnetic field annihilation. Inside
the neutron star the regular toroidal, dipole, or quadrupole poloidal
components of the magnetic field could remain for a long time,
much longer than their chaotic values.

\section*{Acknowledgments}
The authors would like to thank RFBR for their partial support in
the frame of the grant No. 02-02-16900,
NATO for the Collaborative Linkage Grant, the Royal
Society for the grant in the frames of the Joint Projects programme.
We would like to thank Sandra Lilley for significant improvement of the English
language of the paper.




\bsp

\label{lastpage}


\begin{thebibliography}{99}

\bibitem[\protect\citeauthoryear{Akiyama et al.}{2003}]{akiyama}
Akiyama S., Wheeler J.C., Meier D.L., Lichtenstadt I., 2003, ApJ, 584, 954

\bibitem[\protect\citeauthoryear{Ardeljan et al.}{1996}]{abkkm}
Ardeljan N.V.,  Bisnovatyi-Kogan G.S., {\phantom{*}}Kosmachevskii K.V.,
          Moiseenko S.G., 1996, A\&AS,  115, 573

\bibitem[\protect\citeauthoryear{Ardeljan et al.}{2004}]{abkkm2004}
Ardeljan N.V.,  Bisnovatyi-Kogan G.S., {\phantom{*}}Kosmachevskii K.V.,
          Moiseenko S.G., 2004, Astrophysics, 47, 1

\bibitem[\protect\citeauthoryear{Ardeljan et al.}{2000}]{abkm}
Ardeljan N.V., Bisnovatyi-Kogan G.S., Moiseenko S.G.,
  2000, A\&A,  355, 1181

\bibitem[\protect\citeauthoryear{Ardeljan et al.}{1979}]{abkp}
Ardeljan N.V., Bisnovatyi-Kogan G.S., Popov Yu.P.,  1979,
Astron. Zh. 56, 1244

\bibitem[\protect\citeauthoryear{Ardeljan et al.}{1987a}]{abkpch}
Ardeljan N.V., Bisnovatyi-Kogan G.S., Popov Yu.P., Cher\-ni\-gov\-sky S.V., 1987a,
Astron. Zh. 64, 761 (Soviet Astro\-nomy, 1987, 31, 398)

\bibitem[\protect\citeauthoryear{Ardeljan \& Kosmachevskii} {1995}]{arko} Ardeljan N.V,
         Kosmachevskii K.V. 1995, Computational mathematics
         and modeling,  6, 209

\bibitem[\protect\citeauthoryear{Ardeljan et al.}{1987b}]{arkoche} Ardeljan N.V,
Kosmachevskii K.V., Chernigovskii S.V., 1987b, Problems of
         construction and research of conservative difference
         schemes for magneto-gas-dynamics, MSU, Moscow (in
         Russian)

\bibitem[\protect\citeauthoryear{Balbus \& Hawley}{1998}]{balbus}
Balbus S.A., Hawley J.F., 1998, Rev. Mod. Phys., 70, 1

\bibitem[\protect\citeauthoryear{Baym et al.}{1971}]{bps}
Baym G., Pethick C.,  Sutherland P., 1971, ApJ, 170, 299


\bibitem[\protect\citeauthoryear{Bisnovatyi-Kogan}{1970}]{bk1970}
Bisnovatyi-Kogan G.S., 1970,
Astron. Zh. 47, 813 (Soviet Astro\-nomy, 1971, 14, 652)

\bibitem[\protect\citeauthoryear{Bisnovatyi-Kogan et al.}{1976}]{bkpopsam}
Bisnovatyi-Kogan G.S., Popov Yu.P., Samokhin A.A., $\phantom{x}$ 1976, ApSS,  41, 321

\bibitem[\protect\citeauthoryear{Buras et al.}{2003}]{Buras2003}
Buras R., Rampp M., Janka H.Th., Kifonidis K., 2003, Phys.Rev.Lett., 90, 241101

\bibitem[\protect\citeauthoryear{Burrows et al.}{1995}]{Burrows1995}
Burrows A., Hayes J., Fryxell B.A., 1995, ApJ, 450, 830

\bibitem[\protect\citeauthoryear{Chandrasekhar}{1981}]{chandr}
Chandrasekhar S., Hydrodynamic and Hydromagnetic stability New York, Dover,1981

\bibitem[\protect\citeauthoryear{Colgate et al.}{1990}]{colgate}
Colgate S.A., Krauss L.M., Shramm D.N., Walker T.P., 1990, Astro.
Lett. and Coomunications, 27, 411

\bibitem[\protect\citeauthoryear{Cowling}{1957}]{cowling}
Cowling T.G., Magnetohydrodynamics New York, Interscience,1957

\bibitem[\protect\citeauthoryear{Dungey}{1958}]{dungey}
Dungey J.W.,1958, Cosmic electrodynamics. Cambridge Univ. Press, Cambridge

\bibitem[\protect\citeauthoryear{Fromang et al.}{2004}]{fromang}
Fromang S., Balbus S.A., De Villiers J.P., 2004, ApJ, 616, 357


\bibitem[\protect\citeauthoryear{Hawley \& Balbus}{1991}]{hawbalb}
Hawley J.F., Balbus S.A., 1991, ApJ, 376, 223

\bibitem[\protect\citeauthoryear{Ivanova et al.}{1969}]{iin}
Ivanova L.N., Imshennik V.S., Nadezhin D.K., 1969, Na\-uchn. Inform. Astron. Sov.
Akad. Nauk SSSR (Sci. Inf. of the Astr. Council of the
Acad.  Sci.  USSR), 13, 3

\bibitem[\protect\citeauthoryear{Janka \& Plewa}{2002}]{jankaplewa}
Janka H.-Th., Plewa T., astro-ph/0212314

\bibitem[\protect\citeauthoryear{Kosovichev \& Popov}{1979}]{kospop}
Kosovichev A.G., Popov Yu.P., 1979, J. Mathem. Phys. Comput. Math. 19,1251


\bibitem[\protect\citeauthoryear{Kotake et al.}{2004}]{kotake}
Kotake K., Sawai H., Yamada S., Sato K., 2004, ApJ, 608, 391


\bibitem[\protect\citeauthoryear{Lamb}{1909}]{lamb}
Lamb H., 1909, Proc. London math. Soc., 7, 122

\bibitem[\protect\citeauthoryear{Le Blanck \& Wilson}{1970}]{leblanck}
Leblanck L.M., Wilson J.R., 1970, ApJ, 161, 541

\bibitem[\protect\citeauthoryear{Lovelace}{1976}]{lovelace}
Lovelace R.V.E, 1976, Nature, 262, 649

\bibitem[\protect\citeauthoryear{Malone et al.}{1975}]{mjb}
Malone R.C., Johnson M.B.,  Bethe H.A., 1975, ApJ, 199, 741

\bibitem[\protect\citeauthoryear{Meier et al.}{1976}]{meier}
Meier D.L., Epstein R.I., Arnett W.D., \& Schramm D.N. 1976, ApJ, 204, 869

\bibitem[\protect\citeauthoryear{Meier \& Nakamura}{2003}]{meiernakamura}
Meier D.L., Nakamura M. 2003 in Proc. "3-D Signatures in Stellar Explosions",
ed. C.Wheeler, astro-ph/0312050


\bibitem[\protect\citeauthoryear{Ohnishi}{1983}]{ohnishi}
Ohnishu N., 1983, Tech. Rep. Inst. At. En. Kyoto Univ., No.198

\bibitem[\protect\citeauthoryear{Samarskii \& Popov} {1992}]{sampop}
Samarskii A.A., Popov Ju.P., 1992, Difference Methods for the Solution of
Problems of Gas Dynamics. Moscow, Nauka, (in Russian)

\bibitem[\protect\citeauthoryear{Schindler et al.}{1987}]{schindler}
Schinder P.J., Schramm D.N., Wiita P.J., Margolis S.H., Tubbs D.L., 1987, ApJ,
313, 531

\bibitem[\protect\citeauthoryear{Spruit}{2002}]{spruit}
Spruit H.C., 2002, A\&A, 381, 923

\bibitem[\protect\citeauthoryear{Symbalisty}{1984}]{symbalisty}
Symbalisty E.M.D., 1984, ApJ, 285, 729

\bibitem[\protect\citeauthoryear{Tayler}{1973}]{tayler}
Tayler R.J., 1973, MNRAS, 161, 365

\bibitem[\protect\citeauthoryear{Thompson \& Beloborodov}{2004}]{belobor}
Thompson C., beloborodov A.M., 2005, astro-ph/0408538

\bibitem[\protect\citeauthoryear{Velikhov}{1959}]{velikhov}
Velikhov E.P., 1959, J. Exper. Theor. Phys., 36, 1398

\bibitem[\protect\citeauthoryear{Yamada \& Sawai}{2004}]{yamada}
Yamada S., Sawai H., 2004, ApJ, 608, 907

\bibitem[\protect\citeauthoryear{Zenkevich \& Morgan}{1983}]{zemo}
Zenkevich O.C., Morgan K. Finite elements and approximation. NY,
1983, 318с.

\end{thebibliography}
\end{document}